# Structural, elastic, thermal and lattice dynamic properties of new 321 MAX phases

M.A. Hadi,[1*] M.A. Rayhan,[2] S.H. Naqib,[1] A. Chroneos,[3†] A.K.M.A. Islam[1,4]

[1]Department of Physics, University of Rajshahi, Rajshahi 6205, Bangladesh
[2]Department of Arts & Sciences, Bangladesh Army University of Science and Technology, Saidpur 5310, Nilphamari, Bangladesh
[3]Faculty of Engineering, Environment and Computing, Coventry University, Priory Street, Coventry CV1 5FB, UK
[4]International Islamic University Chittagong, Kumira, Chittagong 4318, Bangladesh

**Abstract**

A new series of MAX family designated as 321 phases are recently reported with $Nb_3As_2C$, $V_3As_2C$, $Nb_3P_2C$ and $Ta_3P_2C$. Most of physical properties of these new MAX phase compounds are unexplored and the present study aims to investigate their structural, elastic, thermal and lattice dynamic properties. Their mechanical and dynamical stabilities are examined. Though all the phases are elastically anisotropic and brittle in nature, $Nb_3As_2C$ is the most anisotropic and $Ta_3P_2C$ is the most brittle. $V_3As_2C$ compared to $Nb_3As_2C$ and $Ta_3P_2C$ compared to $Nb_3P_2C$ should exhibit superior mechanical properties, as they possess larger elastic constants and moduli. Shear strength, bond covalency as well as the average bond strength and materials' brittleness are predicted to follow the order: $Ta_3P_2C > Nb_3P_2C > V_3As_2C > Nb_3As_2C$. The estimated Debye temperature and lattice thermal conductivity are highest for $Nb_3As_2C$. Lattice dynamical features are investigated in details and the infrared and Raman active modes are identified. High melting temperatures of these compounds are favorable for their potential applications at elevated temperatures.

## 1. Introduction

MAX phases is a family of layered ternary carbides and nitrides with chemical formula $M_{n+1}AX_n$ ($n$ = 1-3) in which M is an early transition metals, A is a group-A elements and X is carbon or nitrogen [1]. This family of compounds is classified into three main sub-families such as $M_2AX$ or 211 phases, $M_3AX_2$ or 312 phases and $M_4AX_3$ or 413 phases. This family was first discovered in the 1960s with some members of 211 phases known as H-phases [2]. Importantly, in 1996, a phase pure dense $Ti_3SiC_2$ and subsequently a nitride phase $Ti_4AlN_3$ were discovered [3,4] and consequently, a realization has grown up that this family represents a large number of compounds that behave similarly. This realization led to the nomenclature "$M_{n+1}AX_n$" (later abbreviated to MAX) phases for this ternary family.

Experimental studies determined that the MAX phases combine the unusual properties of both metals and ceramics [5–9]. Similar to their matching binary carbides and nitrides (MX), the MAX phases are elastically stiff, thermally and electrically conductive, and resistant to chemical attack and have comparatively low thermal expansion coefficients [10]. Mechanically, they are also very similar. They are reasonably soft and typically readily machinable, damage tolerant and resistant to thermal shock. Likewise, some are creep, fatigue, and oxidation resistant. At ambient temperature, they can be compressed to stresses as high as 1 GPa and fully recover upon taking away of the load, while dissipating approximately 25% of the mechanical energy [11]. At elevated temperatures, they go through a brittle-to-plastic transition (BPT), above which they are quite plastic even under tension [12].

This family of compounds is sometimes termed as nanolaminates because of their laminated layers of thickness in the nanometer range [13]. Layered structure of MAX phases in which $n$ "ceramic" layer(s) sandwiched by an A "metallic" layer is responsible for their ceramic and metallic properties [14]. Due to unusual combination of properties, MAX phases have potential applications as tough, machinable and thermal shock refractories, coatings for electrical contacts, high temperature heating elements, neutron irradiation resistant parts for nuclear applications, precursor for the synthesis of carbide-derived carbon and MXenes, a family of two-dimensional transition metal carbides, nitrides, and carbonitrides [15–22].

---

[*] M.A. Hadi (hadipab@gmail.com)
[†] A. Chroneos (ab8104@coventry.ac.uk)

Very recently, a new series of MAX compounds in the 321 phases has been reported with $Nb_3As_2C$, $V_3As_2C$, $Nb_3P_2C$ and $Ta_3P_2C$ [23]. The new phases differ from the typical MAX phases as they contain an alternate staking of one MX layer and two MA layers in their unit cell, while only one MA layer is allowed in the unit cell of MAX phases. The crystal structure of 321 phases exhibit distinctive structural features from the conventional MAX phases. There are two different M sites in the unit cell of 321 phases, as labeled by M1 and M2 sites similar to 312 phases. Each A atoms is coordinated by three M1 atoms and three M2 atoms forming a triangular prism; each X atom is coordinated by six M1 atoms forming an octahedron. Different from usual MAX phases, the MA triangular-prism layers here are bi-layers. The crystal structure of 321 phases can be visualized as an alternate stacking of MX octahedron layer and MA triangular-prism bi-layer. Similar to MAX phases, the M, A, and X atoms of 321 phases are all in hexagonal closed packing positions. The new phases also crystallize in hexagonal space group $P6_3/mmc$ and can be expressed as $M_{n+1}A_nX$ ($n = 2$) phases.

The reported 321 phases are synthesized experimentally and simultaneously characterized by the theoretical elastic properties and measured bulk modulus only for the $Nb_3As_2C$ phase [23]. Other physical properties of these new compounds are waiting for exploration, which encourage us to perform the present density functional theory (DFT) based first-principles calculations. In this paper, the structural, elastic, and thermal properties and lattice dynamics are investigated systematically.

## 2. Methods of calculations

The present study is employed with the DFT-based first-principles method [24,25] implemented within the Cambridge Serial Total Energy Package (CASTEP) [26]. The electronic exchange-correlation potential is evaluated with the generalized gradient approximation (GGA) [27] within the scheme of Perdew, Burke, and Ernzerhof (PBE) [28]. Vanderbilt type ultrasoft pseudopotential [29] is used to treat the interaction between electron and ion cores. A Γ-centered k-point mesh of 17×17×3 grid in Monkhorst-Pack (MP) scheme [30] is used in the reciprocal space for sampling the first Brillouin zone of hexagonal unit cell for new 321 phases. To expand the eigenfunctions of the valence and nearly valence electrons, a planewave cutoff energy of 540 eV is selected from several trials. Broyden-Flecther-Goldferb-Shinno (BFGS) algorithm is used to minimize the total energy and internal forces [31]. The convergence tolerance is set as follows: difference in total energy less than $5\times10^{-6}$ eV/atom, maximum ionic Hellmann–Feynman force less than 0.01 eV/Å, maximum ionic displacement less than $5\times10^{-4}$ Å, and maximum stress less than 0.02 GPa. Mulliken population analysis is carried out using a projection of the planewave states onto a localized linear combination of atomic orbital (LCAO) basis via a method developed by Sanchez-portal et al. [32]. The population analysis of the resulting projected states is then employed with the Mulliken formalism [33]. The finite-strain method as implemented within CASTEP code is applied to evaluate the elastic properties [34]. The convergence criteria for elastic calculation are: the difference in total energy within $10^{-6}$ eV/atom, and the maximum ionic Helmann-Feynman force within $2\times10^{-3}$ eV/Å, and the maximum ionic displacement less than $10^{-4}$ Å. The lattice dynamical properties such as phonon dispersion, phonon density of states (PHDOS) are calculated using density functional perturbation theory (DFPT) with finite displacement supercell method as implemented in the code [35]. The supercell is defined by the cutoff radius of 5.0 Å, which ensures supercell volume of nine times that of the conventional unit cell for $Nb_3As_2C$, $Nb_3P_2C$ and $Ta_3P_2C$ and 16 times for $V_3As_2C$.

## 3. Result and discussion

### 3.1. Structural properties

The new series of MAX phases designated as 321 phases crystallizes in the hexagonal structure with space group $P6_3/mmc$ (No. 194). The atomic positions are quite different from those of corresponding 312 phases. The atomic positions of M1 and M2 in 321 phases are 4f and 2c, while these atoms in 312 phases occupy the Wyckoff positions of 2a and 4f, respectively. The A-atoms in 321 phases occupy the 4f site but in 312 phases reside in 2b/2d positions. The carbon atoms in 321 phases occupy the 2a Wyckoff positions, while in 312 phases this atom resides in 4f atomic positions. The hexagonal ratio, $c/a$ in the new phases is somewhat smaller than that of 312 phases [36]. The calculated lattice constants are tabulated in Table 1 and shown in Fig. 1. Calculated lattice parameters

*a* and *c* fairly coincide with the experimental values for all phases except *a* for $V_3As_2C$. Among four 321 phases, $Nb_3As_2C$ is phase pure and the present optimized geometry is very close to its experimental geometry. Therefore, the present calculations are deemed to be reliable.

**Table 1**. Geometrical information of new 321 phases; *a* and c are lattice constants, *c/a* is hexagonal ratio, *V* is unit cell volume and $z_M$ and $z_A$ are internal parameters for M1 and A atoms.

| Phases | *a* (Å) | *c* (Å) | *c/a* | *V* (Å$^3$) | $z_M$ | $z_A$ | Reference |
|---|---|---|---|---|---|---|---|
| $Nb_3As_2C$ | 3.3606 | 18.6951 | 5.5631 | 182.85 | 0.0595 | 0.1585 | Expt. [23] |
| | 3.3669 | 18.7946 | 5.5822 | 184.51 | 0.0592 | 0.1582 | This calc. |
| | 3.3857 | 18.9088 | 5.5849 | 187.71 | - - - - - | - - - - - | Calc. [23] |
| $V_3As_2C$ | 3.2985 | 18.0068 | 5.4591 | 169.66 | 0.0634* | 0.1604* | Expt. [20] |
| | 3.1577 | 17.8406 | 5.6499 | 154.05 | 0.0571 | 0.1573 | This calc. |
| | 3.1657 | 17.8618 | 5.6429 | 155.02 | - - - - - | - - - - - | Calc. [23] |
| $Nb_3P_2C$ | 3.3063 | 18.0808 | 5.4686 | 171.15 | 0.0627 | 0.1634 | Expt. [23] |
| | 3.3152 | 18.1058 | 5.4615 | 172.33 | 0.0632 | 0.1604 | This calc. |
| | 3.3334 | 18.2091 | 5.4626 | 175.22 | - - - - - | - - - - - | Calc. [23] |
| $Ta_3P_2C$ | 3.2968 | 18.0101 | 5.4629 | 169.53 | 0.0634 | 0.1626 | Expt. [23] |
| | 3.3494 | 18.2847 | 5.4591 | 177.65 | 0.0644 | 0.1606 | This calc. |
| | 3.3128 | 18.1167 | 5.4687 | 172.19 | - - - - - | - - - - - | Calc. [23] |

*DFT values

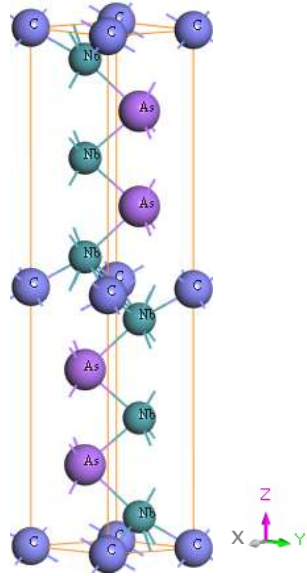

**Fig. 1**. Crystal structure of the 321 MAX phase $Nb_3As_2C$.

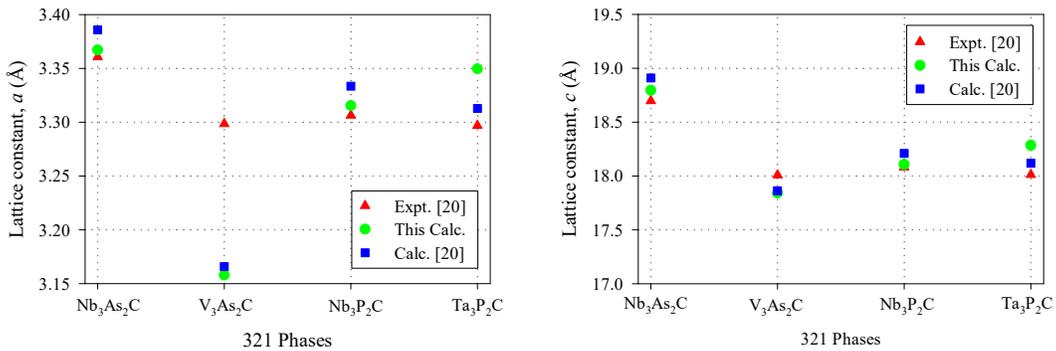

**Fig. 2**. Lattice constants *a* and *c* (in Å) of new 321 MAX phases.

## 3.2. Mechanical properties

Mechanical stability and behaviors of crystalline solids can be assessed with the elastic constants and moduli. The calculated elastic constants $C_{ij}$ are tabulated in Table 2 and shown in Fig. 3 (left). The new 321 MAX phases are mechanically stable as they obey the stability criteria for hexagonal solids [37] with the calculated $C_{ij}$. The investigated elastic moduli are listed in Table 3 and depicted in Fig. 3 (right). In the $M_3As_2C$ (M = Nb, V) and $M_3P_2C$ (M = Nb, Ta) systems, all the elastic constants and moduli calculated in the present study show an increasing trend as Nb is replaced with V and Ta in both systems, respectively. $V_3As_2C$ in $M_3As_2C$ and $Ta_3P_2C$ in $M_3P_2C$ systems should exhibit superior mechanical properties, as they possess larger elastic constants and moduli compared to another member in their own system. Over all, the $M_3P_2C$ phases are mechanically superior to the $M_3As_2C$ phases and $Ta_3P_2C$ is mechanically superior to the other 321 phases. The 211 MAX phases with P as A-element have largest elastic constants, which are consistent for 321 phases studied here [38].

**Table 2**. Calculated elastic constants $C_{ij}$ (in GPa) for the new 321 MAX phases.

| Phases | $C_{11}$ | $C_{12}$ | $C_{13}$ | $C_{33}$ | $C_{44}$ | Reference |
|---|---|---|---|---|---|---|
| $Nb_3As_2C$ | 291.1 | 94.4 | 132.1 | 334.0 | 168.2 | This Calc. |
|  | 290.0 | 120.0 | 150.5 | 362.4 | 179.0 | Calc. [23] |
| $V_3As_2C$ | 301.1 | 101.5 | 146.6 | 352.0 | 178.5 | This Calc. |
|  | 280.4 | 102.6 | 134.3 | 327.6 | 172.9 | Calc. [23] |
| $Nb_3P_2C$ | 337.9 | 103.3 | 155.5 | 396.5 | 193.4 | This Calc. |
|  | 359.9 | 109.3 | 164.7 | 415.2 | 201.3 | Calc. [23] |
| $Ta_3P_2C$ | 353.5 | 106.3 | 164.8 | 414.4 | 198.6 | This Calc. |
|  | 350.6 | 105.1 | 167.0 | 418.6 | 197.7 | Calc. [23] |

**Table 3**. Calculated bulk, shear and Young's moduli ($B_V$, $G_V$, $E_V$ in GPa) and Pugh's and Poisson's ratios ($P_V$, $\sigma_V$) for the new 321 MAX phases.

| Phases | $B_V$ | $G_V$ | $E_V$ | $P_V$ | $\sigma_V$ | Reference |
|---|---|---|---|---|---|---|
| $Nb_3As_2C$ | 180.8 | 124.6 | 304.0 | 1.45 | 0.22 | This Calc. |
|  | 198.3 | 123.4 | 306.5 | 1.61 | 0.24 | Calc. [23] |
| $V_3As_2C$ | 193.7 | 128.7 | 316.1 | 1.51 | 0.23 | This Calc. |
|  | 181.2 | 121.4 | 297.8 | 1.49 | 0.23 | Calc. [23] |
| $Nb_3P_2C$ | 211.2 | 144.7 | 353.4 | 1.46 | 0.22 | This Calc. |
|  | 223.6 | 152.0 | 371.8 | 1.47 | 0.22 | Calc. [23] |
| $Ta_3P_2C$ | 221.4 | 149.9 | 366.9 | 1.48 | 0.22 | This Calc. |
|  | 222.0 | 149.0 | 365.3 | 1.49 | 0.23 | Calc. [23] |

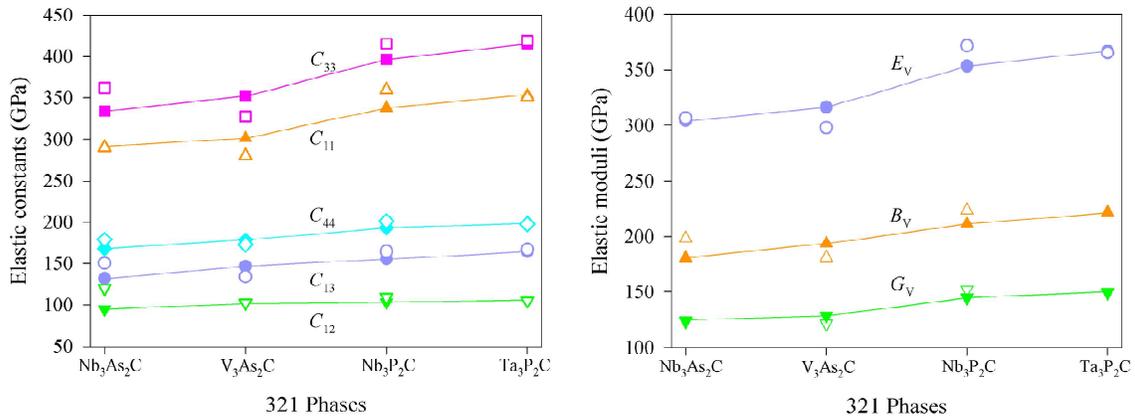

**Fig. 3**. Elastic constants and moduli of 321 MAX phases. The filled and open symbols stand for the present and literature values [23], respectively.

For all phases it is observed that $C_{33} > C_{11}$, indicating that the new phases are more incompressible along the c-axis than the a-axis. It also indicates that the elastic anisotropy in the 321 MAX phases is profound. The shear elastic constants $C_{12}$ and $C_{13}$ lead together to a functional stress component in the crystallographic a-axis with a uniaxial strain along the crystallographic b- and c-axis, respectively. The comparatively large values of these two constants of 321 phases make them capable of resisting the shear deformation along the crystallographic b- and c-axis under a large stress along the crystallographic a-axis. The rank of shear strength of four 321 phases can be shown as, $Ta_3P_2C > Nb_3P_2C > V_3As_2C > Nb_3As_2C$.

The difference ($C_{12} - C_{44}$) is known as Cauchy pressure and can serve as an indicator of brittle/ductile failure of solids. A brittle (ductile) material always has negative (positive) Cauchy pressure. Therefore, all members of new 321 phases are brittle in nature. The nature of chemical bonding in crystalline solids can be predicted with Cauchy pressure. The positive Cauchy pressure is indicative of metallic bonding, while negative Cauchy pressure signifies the directional covalent bonding with angular character. All the 321 phases are dominated with directional covalent bonding as they have negative Cauchy pressure and follow the same order of shear strength for their covalency and brittleness. It may be concluded that the higher the shear strength the stronger the covalent bonding.

The elastic constant $C_{44}$ indirectly leads to the indentation hardness of the materials. Due to comparatively large $C_{44}$, the hardness of 321 phases may exceed the limit of 2-8 GPa for most of the MAX phases. A high $C_{44}$ is indicative of low shearability and low machinability of the compounds. Therefore, the new 321 phases are hard materials compared to conventional MAX phases.

The calculated bulk modulus $B_V$, shear modulus $G_V$, Young's modulus $E_V$ [$9B_VG_V/(3B_V + G_V)$], Pugh's ratio $P_V$ ($B_V/G_V$) and Poisson's ratio $\sigma_V$ [$(3B_V - 2G_V)/(6B_V + 2G_V)$] are listed in Table 3. The bulk modulus provides a measure of resistance to volume change and the average bond strength in a material. The higher the bulk modulus the stronger is the bond strength. The average bond strength should follow this order: $Ta_3P_2C > Nb_3P_2C > V_3As_2C > Nb_3As_2C$ same as the order of shear strength and bond covalency. Shear modulus is another important physical quantity, which assesses the ability to resist the materials' shape change. Higher shear modulus leads to difficult shape change. The resistance to shape change also follows the same order of average of bond strength, shear strength and bond covalency. The bulk modulus is seen to be always greater than the shear modulus, indicating that the factor limiting the mechanical stability is of the 321 phases is the shear modulus.

Young's modulus usually is used to describe the resistance of a material against uniaxial tension and stiffness. High Young's modulus of 321 phases indicates high stiffness of these materials. The rank of stiffness can be described as the rank of average of bond strength. Pugh's ratio defined as bulk to shear modulus ratio can classify the crystalline solids as brittle or ductile materials with a decisive value of 1.75 [39]. The brittle materials have values lower than this value, while the ductile materials have values greater than this value. According to this decisive factor the studied 321 MAX phases are brittle in nature as found with Cauchy pressure.

Poisson's ratio ($\sigma_V$) can assess many physical properties of solids. Stability against shear can be predicted with the Poisson's ratio. Lower Poisson's ratio leads to stability of the compounds against shear [40]. With their low Poisson's ratios the new 321 MAX phases are expected to be stable against shear. Poisson's ratio can also predict the nature of interatomic forces in the crystalline solids. Poisson's ratio ranges from 0.25 to 0.50 for central force solids and lies outside this range for non-central force solids [41]. Evidently, all the new compounds belong to the latter group. Failure mode of solids can also be justified with the Poisson's ratio. A value of 0.26 can serve as the division line between the brittle and ductile materials [42,43]. A brittle material lies under this line and a ductile material exceeds this line. Accordingly, all the phases in the 321 MAX family are brittle in nature. Poisson's ratio plays another important role in assessing the nature of chemical bonding in solid materials [44]. The value of Poisson's ratio for pure covalent crystal is 0.1. Conversely, the completely metallic compounds possess a value of 0.33. The Poisson's ratio for four members of new series of MAX phases lies between these two characteristic values. Consequently, it is expected that the chemical bonding in new compounds is a mixture of covalent and metallic in nature.

All the studied compounds are elastically anisotropic. There are different types of parameters for evaluating the level of elastic anisotropy in different views. There are three shear anisotropy factors for hexagonal crystals like studied compounds associated with three different shear planes. The expressions for these factors are as follows [45]:

$$A_1 = \frac{(C_{11} + C_{12} + 2C_{33} - 4C_{13})}{6C_{44}}$$

for {100} shear planes between ⟨011⟩ and ⟨010⟩ directions.

$$A_2 = \frac{2C_{44}}{C_{11} - C_{12}}$$

for {010} shear planes between ⟨101⟩ and ⟨001⟩ directions.

$$A_3 = \frac{(C_{11} + C_{12} + 2C_{33} - 4C_{13})}{3(C_{11} - C_{12})}$$

for {001} shear planes between ⟨110⟩ and ⟨010⟩ directions.

Shear anisotropy factors $A_i$ ($i$ = 1, 2, 3) have unit value for isotropic crystals. Deviation of $A_i$ from unity ($\Delta A_i$) evaluates the degree of elastic anisotropy in shear. The calculated $\Delta A_i$ is listed in Table 4 and shown in Fig. 4. As seen from Table 4 as well as from Fig. 4, the new series of MAX phases exhibit profound anisotropy on the {1 0 0} and {0 1 0} shear planes, while on the {1 0 0} plane the anisotropy is very small. The ratio of linear compressibility coefficient along a- and c-axis, $k_c/k_a$ provides a measure of elastic anisotropy for hexagonal crystal [45]. This factor is defined as:

$$k_c/k_a = \frac{(C_{11} + C_{12} - 2C_{13})}{(C_{33} - C_{13})}$$

The unit vector of this ratio indicates the isotropic nature of the crystals. Deviation from unity $\Delta(k_c/k_a)$ assesses the level of anisotropy. The calculated $\Delta(k_c/k_a)$ is listed in Table 4 and depicted in Fig. 4, indicating that the compressibility along the c-axis is smaller than that along the a-axis for all 321 phases. A universal factor for quantifying the elastic anisotropy for all classes of crystals is introduced by Ranganathan and Ostoja-Starzewsky [46]. This factor is known as universal anisotropy index and defined as:

$$A^U = 5\frac{G_V}{G_R} + \frac{B_V}{B_R} - 6$$

This factor has either zero or positive value. The zero value is indicative of purely isotropic crystals, while the positive value signifies the anisotropy level in crystalline solids. The calculated values are listed in Table 4 and illustrated in Fig. 4, indicating significant anisotropy of the newly synthesized compounds. The difference between Voigt and Reuss bulk modulus ($B_V$, $B_R$) and shear modulus ($G_V$, $G_R$) leads to two anisotropy factors known as percentage anisotropy factors $A_B$ and $A_G$ [42]:

$$A_B = \frac{B_V - B_R}{B_V + B_R} \times 100\%$$

$$A_G = \frac{G_V - G_R}{G_V + G_R} \times 100\%$$

**Table 4**. Elastic anisotropy factors calculated for new series of MAX phases.

| Phases | $\Delta A_1$ | $\Delta A_2$ | $\Delta A_3$ | $\Delta(k_c/k_a)$ | $A^U$ | $A_B$ | $A_G$ | Reference |
|---|---|---|---|---|---|---|---|---|
| Nb$_3$As$_2$C | 0.4797 | 0.7102 | 0.1102 | 0.3992 | 0.4513 | 0.8288 | 4.1646 | This Calc. |
|  | 0.5039 | 1.1059 | 0.0447 | 0.4856 | 0.7245 | 1.1263 | 6.5573 | Calc. [23]* |
| V$_3$As$_2$C | 0.5143 | 0.7886 | 0.1313 | 0.4674 | 0.5550 | 1.0270 | 5.0718 | This Calc. |
|  | 0.5171 | 0.9449 | 0.0607 | 0.4082 | 0.6282 | 0.7678 | 5.7737 | Calc. [23]* |
| Nb$_3$P$_2$C | 0.4724 | 0.6488 | 0.1302 | 0.4598 | 0.4329 | 1.0649 | 3.9509 | This Calc. |
|  | 0.4694 | 0.6065 | 0.1476 | 0.4419 | 0.4065 | 0.9594 | 3.7273 | Calc. [23]* |
| Ta$_3$P$_2$C | 0.4718 | 0.6068 | 0.1513 | 0.4784 | 0.4190 | 1.1504 | 3.8068 | This Calc. |
|  | 0.4732 | 0.6106 | 0.1515 | 0.5163 | 0.4335 | 1.3700 | 3.8995 | Calc. [23]* |

*Calculated with published data.

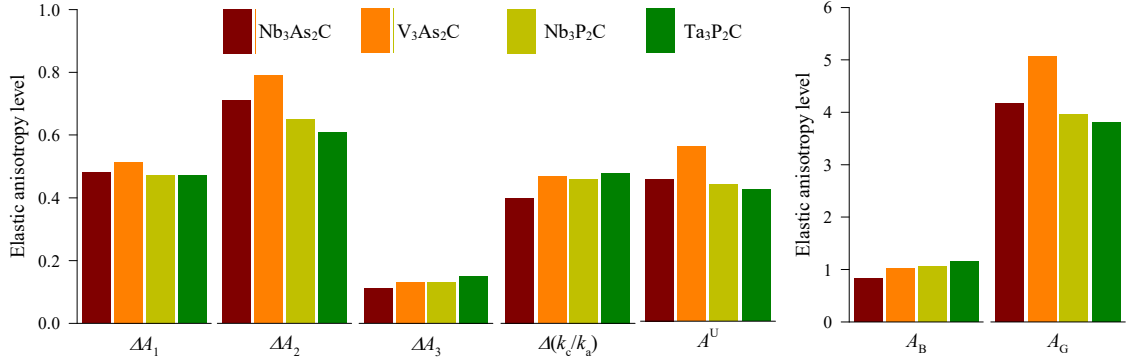

**Fig. 4**. Elastic anisotropy factors for new series of MAX phases.

The calculated percentage anisotropy factors $A_B$ and $A_G$ are also listed in Table 4 and shown in Fig. 4. These two factors have zero values for completely isotropic crystals in view of compressibility and shear, respectively. It is evident that the percentage anisotropy in shear is significant for all 321 phases compared to the percentage anisotropy in compressibility. Among four phases, $Nb_3As_2C$ is elastically more anisotropic in view of most indicators discussed here.

### 3.3. Thermal properties

Debye temperature is a characteristic temperature of crystalline solids that correlates with a range of physical properties, such as thermal conductivity, thermal expansion, lattice vibration, specific heats, melting point, etc. For superconducting materials, Debye temperature makes a bridge between electron-phonon coupling constant and superconducting transition temperature. Besides, it is related to the vacancy formation energy in metals. There are several methods for calculating the Debye temperature. Among them, Anderson method is rigorous and straightforward and depends on average elastic (sound) wave velocity via the following equation [47]:

$$\theta_D = \frac{h}{k_B}\left[\left(\frac{3n}{4\pi}\right)\frac{N_A \rho}{M}\right]^{1/3} v_m$$

where $h$ and $k_B$ is Planck's and Boltzmann's constants, respectively, $N_A$ is the Avogadro's number, $\rho$ is the mass density, M is the molecular weight, and n is the number of atoms in a molecule; $v_m$ is the average sound velocity in the crystalline solids that can be calculated from:

$$v_m = \left[\frac{1}{3}\left(\frac{1}{v_l^3} + \frac{2}{v_t^3}\right)\right]^{-1/3}$$

Here $v_l$ and $v_t$ are the longitudinal and transverse sound velocities, respectively in crystalline solids. These velocities can be calculated using the following equations:

$$v_l = \left[\frac{3B_V + 4G_V}{3\rho}\right]^{1/2} \quad \text{and} \quad v_t = \left[\frac{G_V}{\rho}\right]^{1/2}$$

The calculated Debye temperature and sound velocities are listed in Table 5 and shown in Fig. 5. The Debye temperature is highest for $Nb_3P_2C$ phase and lowest for $Ta_3P_2C$ phase. Higher Debye temperature corresponds to the better thermal conductivity of the compounds. Therefore, $Nb_3P_2C$ is thermally more conductive in the new series of 321 phases.

Fine *et al.* [48] developed an empirical formula for calculating the melting temperature of hexagonal crystal using elastic constants:

$T_m = 354 + 1.5(2C_{11} + C_{33})$

The calculated melting temperatures of 321 phases is listed in Table 5 and shown in Fig. 5. The high melting temperatures of these compounds is encouraging for their application as high temperature structural materials.

**Table 5.** The elastic wave velocity, Debye temperature, melting temperature, minimum and lattice thermal conductivity of 321 phases.

| Phases | $\rho$ | $v_t$ | $v_l$ | $v_m$ | $\theta_D$ | $T_m$ | $\kappa_{min}$ | $\kappa_{ph}$** | Reference |
|---|---|---|---|---|---|---|---|---|---|
| Nb$_3$As$_2$C | 7.929 | 3.964 | 6.615 | 4.386 | 525.1 | 1728.3 | 0.98 | 29.34 | This Calc. |
|  | 7.794 | 3.979 | 6.823 | 4.413 | 525.4 | 1767.6 | 0.97 | 25.82 | Calc. [23]* |
| V$_3$As$_2$C | 6.783 | 4.356 | 7.338 | 4.824 | 613.3 | 1785.3 | 1.22 | 29.40 | This Calc. |
|  | 6.741 | 4.244 | 7.134 | 4.698 | 596.1 | 1686.6 | 1.18 | 27.06 | Calc. [23]* |
| Nb$_3$P$_2$C | 6.796 | 4.614 | 7.711 | 5.106 | 625.3 | 1962.5 | 1.19 | 38.78 | This Calc. |
|  | 6.684 | 4.769 | 7.986 | 5.278 | 642.8 | 2056.5 | 1.22 | 42.36 | Calc. [23]* |
| Ta$_3$P$_2$C | 11.530 | 3.606 | 6.045 | 3.991 | 483.9 | 2036.1 | 0.91 | 31.74 | This Calc. |
|  | 11.895 | 3.539 | 5.947 | 3.918 | 480.0 | 2033.7 | 0.92 | 28.68 | Calc. [23]* |

*Calculated with published data.
**Calculated at 300 K.

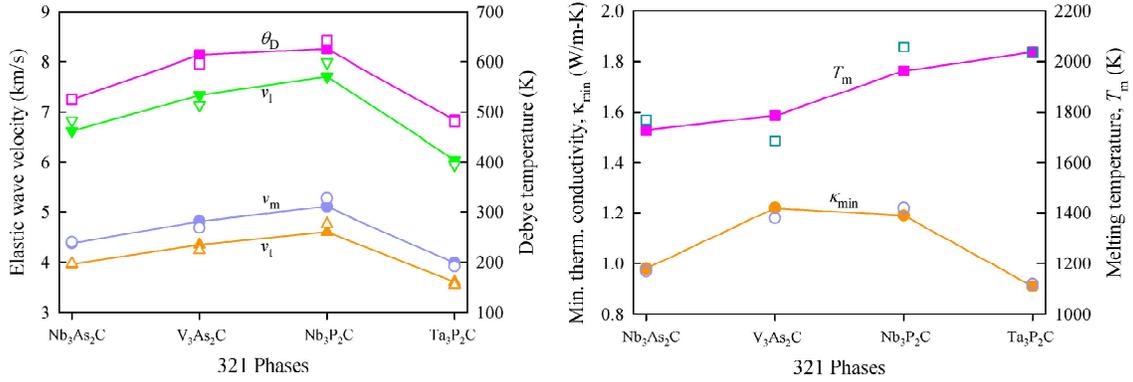

**Fig. 5.** Elastic wave velocities and Debye temperature (left) and minimum thermal conductivity and melting temperature (right) of 321 phases. The open symbols represent the results calculated from literature values [23].

Thermal conductivity determines how the atoms behave within a crystal when it is heated or cooled. Minimum thermal conductivity signifies the theoretical lower limit of intrinsic thermal conductivity of a crystal. It can be calculated from the average sound velocity using [49]:

$$\kappa_{\min} = k_B v_m \left(\frac{nN_A\rho}{M}\right)^{2/3}$$

The calculated minimum thermal conductivity of new series of 321 MAX phases are tabulated in Table 5 and shown in Fig. 5. The minimum thermal conductivity is directly proportional to the average sound velocity in crystalline solids. As the average sound velocity of Ta$_3$P$_2$C is lowest the minimum thermal conductivity of this phase is also lowest.

The lattice thermal conductivity is one of the most useful physical properties of crystals, mainly for their applications at elevated temperatures. DFT calculation of lattice thermal conductivity of MAX phases is very complicated because of their dual properties of metals and ceramics. There are different models for calculating the lattice thermal conductivity of solids. Among them, Slack's model is most efficient in determining this property for ceramic like MAX phases. According to Slack's model, the temperature dependent lattice thermal conductivity $\kappa_{ph}$ of 321 MAX phases can be evaluated using the empirical formula [50]:

$$\kappa_{ph} = A \frac{M_{av}\theta_D^3\delta}{\gamma^2 n^{2/3} T}$$

Here $M_{av}$ is the average atomic mass (in kg/mol) in a molecule, $\theta_D$ is the Debye temperature (in K), $\delta$ is the cubic root of average atomic volume, $n$ is number of atoms per unit cell, $T$ is the absolute temperature, $\gamma$ is the Grüneisen parameter derived from Poison's ratio ($v$) and $A$ is a factor (in W-mol/kg/m$^2$/K$^3$) depending on $\gamma$. The parameter $\gamma$ can be calculated as:

$$\gamma = \frac{9(v_l^2 - (4/3)v_t^2)}{2(v_l^2 - 2v_t^2)} = \frac{3(1+v)}{2(2-3v)}$$

The factor $A(\gamma)$ according to Julian [51] can be calculated as:

$$A(\gamma) = \frac{5.720 \times 10^7 \times 0.849}{2 \times (1 - 0.514/\gamma + 0.228/\gamma^2)}$$

The calculated lattice thermal conductivity of 321 MAX phases at the room temperature (300 K), are given in Table 5 and its temperature dependence is depicted in Fig. 6. The total thermal conductivity of MAX phases at 300 K varies from 12 to 60 W/m-K [52]. Though the electronic contribution, which is very small, is excluded in the present calculations, the present results lie within this range fairly. It is observed that lattice thermal conductivity decreases gradually with the increase of temperature. In the whole range of temperature, the lattice thermal conductivity is highest for $Nb_3P_2C$ and lowest for $Nb_3As_2C$.

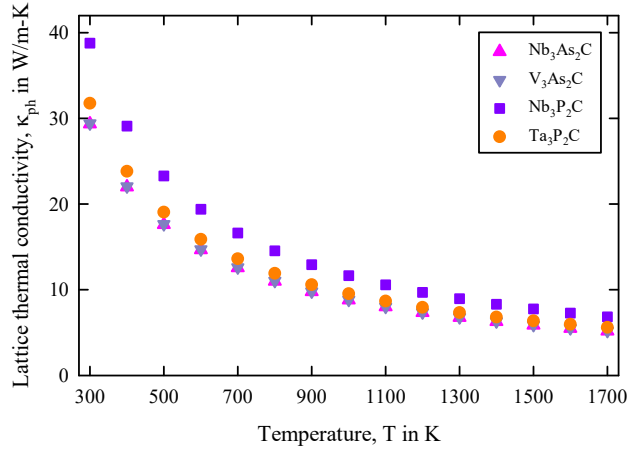

**Fig. 6**. Lattice thermal conductivity of new series of 321 MAX phases.

*3.4. Lattice dynamics*

The study of lattice dynamics mainly includes phonon dispersion and phonon density of states (PHDOS). Dynamical stability of a compound can be verified with these two properties. Phonon dispersion of a new series of 321 MAX phases are calculated along the high symmetry directions at the equilibrium volume and depicted in Fig. 7. In the whole Brillouin zone, there is no sign for negative frequency, indicating the dynamical stability of newly synthesized 321 phases. This 'intrinsic stability' is a fundamental issue for a compound to be a stable one. The lower branches in the dispersion curves are the acoustic branches, which originate as a result of coherent movements of atoms of the lattice outside of their equilibrium positions. The acoustic modes at the Γ point have zero frequency. This is also a sign of dynamical stability of the studied four compounds.

The upper branches represent the optical branches. The optical properties of crystals are mainly controlled by the optical branches, which are due to the out-of-phase movements of atoms in a lattice, one atom moving to the left, and its neighbor to the right. They are named "optical' because in ionic crystals, the phonons associated with these branches are excited by infrared radiation, which leads to a vibrational state in which both positive and negative ions oscillate out-of-phase with respect to each other. Optical phonons have a non-zero frequency at the Brillouin zone center (Γ) and show flat-type band (no dispersion) near that long wavelength limit. This flatness of the optical modes causes a very sharp peak in the phonon density of states. This is because they characterize a mode of vibration where positive and negative ions at adjoining lattice sites swing against each other, generating a time-varying electrical dipole moment.

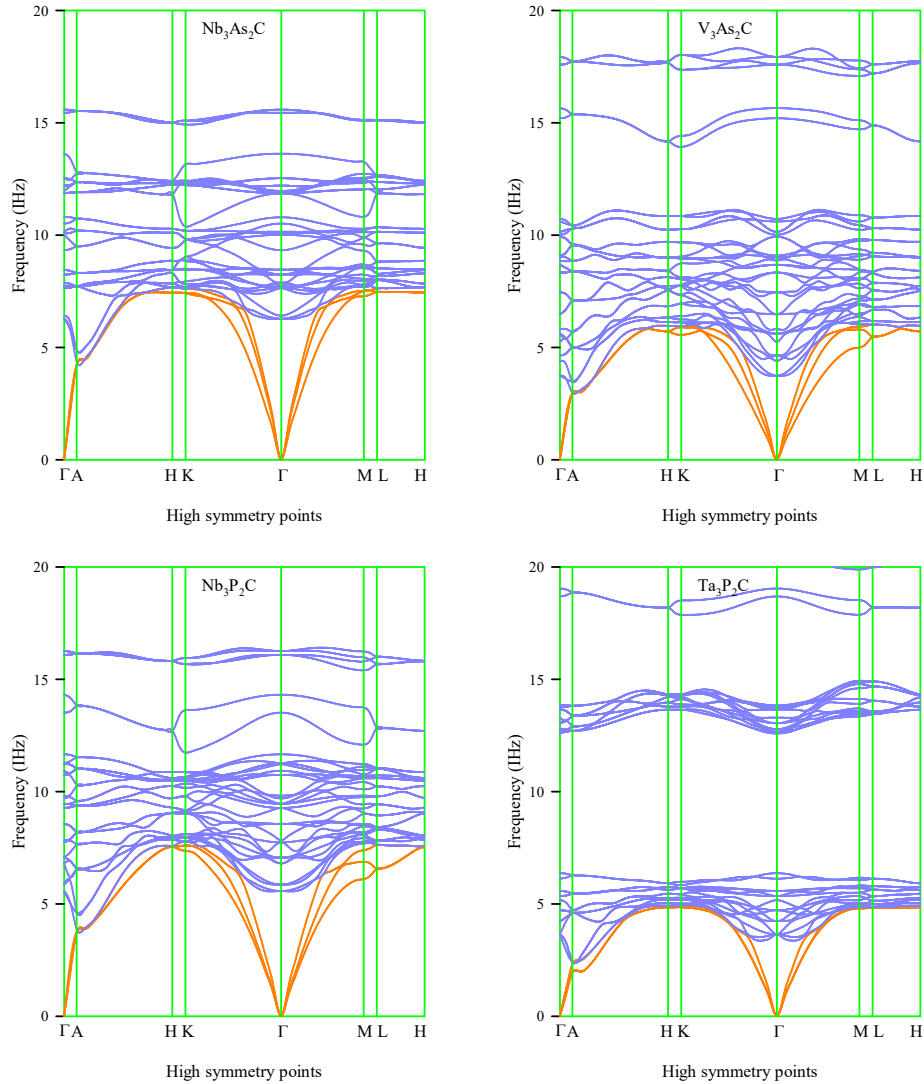

**Fig. 7**. Phonon dispersion curves of new series of 321 MAX phases.

In the $M_3As_2C$ phases, the structures exhibit different features in their phonon band structures, particularly for the higher-frequency optical branches above 14 THz. This is because those branches characterize the vibrational modes of the carbon atom, which have the different environment in both phases (i.e. within $Nb_3C_2$ and $V_3C_2$ octahedra). The similar results arise in the in the phonon band structures for $M_3P_2C$ phases due to same reasons.

The total PHDOS calculated for 321 MAX phases is depicted in Fig. 8. The main contribution to the acoustic branches that confined to an area of the low frequencies results from the transition metal atoms, while the high frequency phonons referring to the optical branches is due to the carbon atoms. This is expected as the carbon atom is lighter than the transition metal atoms, which leads to fairly weaker electron-phonon interactions. A-atoms (As/P) contribute to both the acoustic and optical branches. The highest peak in the optical branches results in vibration of A-atoms. In the $M_3As_2C$ systems, the substitution of Nb by V causes a significant change in shape and position of both the acoustic and optical branches. As a result of this substitution, the acoustic branches shifts to more low frequency region and the optical branches expand to higher frequency region in $V_3As_2C$. However, the upper optical branch in $V_3As_2C$ splits into two segments. In the $M_3P_2C$ systems, the substitution of Nb by Ta results in a large change in their PHDOS profiles as well as in dispersion curves. The acoustic branches condense to low frequency region and large band gaps arise among optical branches.

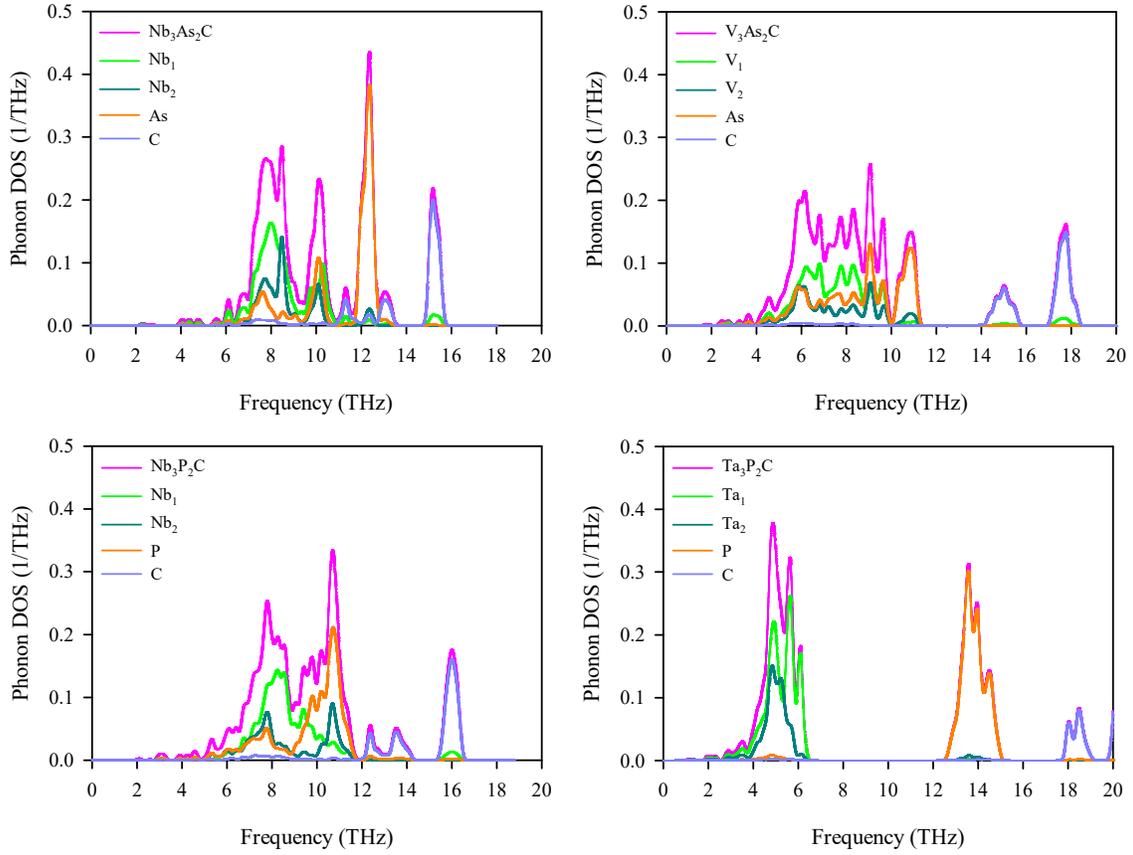

**Fig. 8.** Phonon density of states (PHDOS) for a new series of MAX phases

**Table 6**. Theoretical wavenumbers ω and symmetry assignment of the Infrared and Raman active modes of the 321 phases.

| Mode | Irr. Rep | IR active modes Wavenumbers ω (cm$^{-1}$) | | | | Mode | Irr. Rep | Raman active modes Wavenumbers ω (cm$^{-1}$) | | | |
|---|---|---|---|---|---|---|---|---|---|---|---|
| | | Nb$_3$As$_2$C | V$_3$As$_2$C | Nb$_3$P$_2$C | Ta$_3$P$_2$C | | | Nb$_3$As$_2$C | V$_3$As$_2$C | Nb$_3$P$_2$C | Ta$_3$P$_2$C |
| ω$_1$ | $E_{1u}$ | 255 | 187 | 236 | 157 | ω$_1$ | $E_{2g}$ | 257 | 125 | 186 | 141 |
| ω$_2$ | $A_{2u}$ | 263 | 194 | 261 | 172 | ω$_2$ | $E_{1g}$ | 275 | 216 | 258 | 178 |
| ω$_3$ | $A_{2u}$ | 336 | 334 | 323 | 420 | ω$_3$ | $E_{2g}$ | 282 | 278 | 286 | 186 |
| ω$_4$ | $A_{2u}$ | 399 | 507 | 451 | 623 | ω$_4$ | $E_{1g}$ | 396 | 289 | 316 | 204 |
| ω$_5$ | $E_{1u}$ | 407 | 301 | 358 | 435 | ω$_5$ | $A_{1g}$ | 311 | 303 | 327 | 458 |
| ω$_6$ | $E_{1u}$ | 520 | 587 | 537 | 683 | ω$_6$ | $E_{2g}$ | 418 | 331 | 375 | 455 |
| | | | | | | ω$_7$ | $A_{1g}$ | 360 | 338 | 389 | 423 |

The wavenumbers (equivalent to energy and frequency) of Infrared (IR) active and Raman-active modes for each irreducible representation involved in the 321 phases are listed in Table 6. IR activity modes are related to non-zero transition dipole moment, while Raman activity modes correlate with non-zero transition polarizability. The 12 atoms of the primitive cell of 321 MAX phases produce 36 zone centre vibrational modes. Among them 33 are optical modes and 3 are acoustic modes. The factor group analysis suggests that the irreducible representations for the optical modes of Nb$_3$As$_2$C phase at the Γ point can be expressed as [53]:

$\Gamma_{opt} = 3A_{2u} + 6E_{1u} + 6E_{2u} + 3B_{2g} + 6E_{2g} + 3B_{1u} + 4E_{1g} + 2A_{1g}$

where $E_{1u}$ and $A_{2u}$ are IR active modes and $A_{1g}$, $E_{1g}$ and $A_{2g}$ are Raman active modes. The remaining other modes ($B_{1u}$ and $B_{2g}$) are silent. The 321 phase Nb$_3$As$_2$C has a total of seven Raman active modes ($2A_{1g} + 2E_{1g} + 3E_{2g}$), which is consistence to the 312 phases that have also a total of seven Raman active modes ($2A_{1g} + 2E_{1g} + 3E_{2g}$) [54]. Other 321 phases V$_3$As$_2$C, Nb$_3$P$_2$C, and Ta$_3$P$_2$C have the same results.

A vibrational mode will be Raman-active if its displacements modify the molecular polarization potential, and the relevant Raman intensity depends on the polarizability change induced by the mode. The Raman-active modes in a crystal are a subset of the Γ-point normal modes that governed by the selection-rules and controlled by symmetry. Raman active modes $A_{1g}$, $E_{1g}$ and $A_{2g}$ involve the simultaneous stretching and then simultaneous compression of bonds, which will lead to a distortion in the electron clouds and induce non-zero polarizability. On the other hand, the $E_{1u}$ and $A_{2u}$ bonds are Infrared-(IR) active because IR spectroscopy is governed by different selection rules. As these two modes are IR active they involve changes in dipole moment and as a result, optical absorption arises from the oscillations.

## 4. Conclusions

Using DFT-based first-principles calculations, we explore a set of physical properties of the new 321 MAX family phases, namely $Nb_3As_2C$, $V_3As_2C$, $Nb_3P_2C$ and $Ta_3P_2C$. These new compounds are mechanically and dynamically stable. Elastic anisotropy should be high in $Nb_3As_2C$ and $Ta_3P_2C$ should exhibit highest brittleness. Larger elastic constants and moduli suggest that the $M_3P_2C$ phases are mechanically superior to the $M_3As_2C$ phases and $Ta_3P_2C$ to the other 321 phases. The order of $Ta_3P_2C > Nb_3P_2C > V_3As_2C > Nb_3As_2C$ is seen to be applicable for shear strength, bond covalency, average bond strength and brittleness. The phase $Nb_3As_2C$ possesses the highest values for Debye temperature and lattice thermal conductivity. All the four compounds under study possess quite high melting temperatures and these new compounds are expected to have potential applications at high temperature environments. Lattice dynamical calculations show clear evidences of acoustical and optical branches with varying degree of gaps. We have identified the IR and Raman active vibrational modes of $Nb_3As_2C$, $V_3As_2C$, $Nb_3P_2C$ and $Ta_3P_2C$ which are consistent with the structure and crystal symmetry of the 321 phases under consideration. We hope our study will stimulate the MAX phase research community to explore the characteristics of these compounds both experimentally and theoretically in near future.